\documentclass{aastex63}

\usepackage{CJKutf8}
\usepackage[caption=false]{subfig}
\usepackage{graphicx}

\usepackage{multirow}
\usepackage{graphicx}
\usepackage{booktabs}
\usepackage{amsmath,amssymb}
\usepackage[T1]{fontenc}
\graphicspath{{./}{figures/}}

\begin{document}

\title{iPTF~16asu Revisited: A Rapidly Evolving Superluminous Broad-Lined Ic Supernova?}

\correspondingauthor{Shan-Qin Wang \begin{CJK*}{UTF8}{gbsn}(王善钦)\end{CJK*}}
\email{shanqinwang@gxu.edu.cn}

\author{Shan-Qin Wang \begin{CJK*}{UTF8}{gbsn}(王善钦)\end{CJK*}}
\affiliation{Guangxi Key Laboratory for Relativistic Astrophysics,
School of Physical Science and Technology, Guangxi University, Nanning 530004,
China}

\author{Wen-Pei Gan \begin{CJK*}{UTF8}{gbsn}(甘文沛)\end{CJK*}}
\affiliation{Guangxi Key Laboratory for Relativistic Astrophysics,
School of Physical Science and Technology, Guangxi University, Nanning 530004,
China}

\begin{abstract}

In this paper, we fit the spectral energy distributions (SEDs) of iPTF~16asu
that has so far been classified as a luminous rapidly evolving
broad-lined Ic supernova (SN Ic-BL),
and re-construct its post-peak bolometric light curve. We find that
the luminosity of the post-peak bolometric light curve of iPTF~16asu is
about 3 times that of the pseudo-bolometric light curve derived in the literature, 
and the extrapolated peak luminosity exceeds
$\sim 10^{44}$ erg s$^{-1}$, which is higher than the threshold of
superluminous SNe (SLSNe). We then use the $^{56}$Ni model and the
magnetar plus $^{56}$Ni model to fit the multi-band light curves
of iPTF~16asu, and construct the theoretical bolometric light curve using the
best-fitting theoretical multi-band light curves. We find that the
magnetar plus $^{56}$Ni model can account for the photometry of iPTF~16asu,
and the peak luminosity of its theoretical bolometric light curve
is $\sim 1.06\times 10^{44}$ erg s$^{-1}$.
We suggest that iPTF~16asu and similar SNe (e.g., SN~2018gep)
constitute the class of rapidly evolving SLSNe Ic-BL.

\end{abstract}


\section{Introduction}
\label{sec:intro}

In the past two decades, wide-field optical survey telescopes have discovered
hundreds of extreme optical transients which can be split into two
classes: superluminous supernovae (SLSNe) with peak luminosities
at least 10 times higher than those of ordinary SNe (see
\citealt{Gal2012,Gal2018} for reviews), and rapidly evolving optical
transients (REOTs, e.g., \citealt{Drout2014,Arc2016,Tanaka2016,Pur2018})
whose rise time is $\sim 2-12$ days.
Although the nature of the majority of SLSNe and REOTs is illusive, a major
fraction of SLSNe are believed to be SNe Ic or IIn powered by a magnetar
or ejecta-circumstellar medium (CSM) interaction, while some REOTs have 
been confirmed to be type Ibn and Ic SNe \citep{Ho2021}.

iPTF~16asu discovered by the intermediate Palomar Transient Factory 
(iPTF, \citealt{Cao2016}) on 2016 May 11.26 UT is one of such rapidly evolving
SN Ic at redshift $z$ = 0.187. 
The follow-up observations and analysis show that
the $g-$band light curve of iPTF~16asu peaked at absolute magnitude of $-$20.4 mag
\citep{White2017} (W17 hereafter) which is comparable to $g-$band peaks of SLSNe
PTF10hgi ($M_g, {\rm peak} = -20.42$ mag)
and PTF11rks ($M_g, {\rm peak} = -20.76$ mag) \citep{Inse2013}.
W17 find that late-time spectra of iPTF16asu show the features of the
spectra of broad-lined Type Ic SNe (SNe Ic-BL).

W17 use several models to fit the pseudo-bolometric light curve they constructed.
They find that the full pseudo-bolometric light curve cannot be fitted by the
$^{56}$Ni model and the shock-breakout model, but can be fitted by the magnetar model.
W17 suggest, however, that the derived ejecta mass (0.086\,M$_\odot$) is
too small and conclude that the value is unreasonable. 

As pointed out by W17, the bolometric luminosity of iPTF~16asu might be
underestimated since they constructed the post-peak pseudo-bolometric light
curve by accounting for the observed flux. Using this method, the
peak of the pseudo-bolometric light curve is $(3.4\pm 0.3)\times 10^{43}$ erg s$^{-1}$,
which is lower than the blackbody luminosity at day $+3.47$
($(6.4\pm 1.6)\times 10^{43}$ erg s$^{-1}$) they derived by using the blackbody
model. \footnote{\cite{Wang2019} construct the bolometric light curve of iPTF~16asu
and suggest that its peak luminosity is $3.8\times 10^{43}$ erg s$^{-1}$, which is
slightly higher than that given by W17.}
The pre-peak pseudo-bolometric light curve might also be not precise, since
it is obtained by assuming that the $g-$band flux to total flux is a constant,
while this ratio might not be a constant.

Due to the fact that rapidly evolving (super-)luminous SNe Ic-BL are extremely
rare, the real bolometric light curve and the energy sources of
iPTF~16asu deserve further study. In this paper, we re-investigate these two issues.
In Section \ref{sec:SED}, we re-investigate the SEDs
of iPTF~16asu and derive the bolometric luminosity at some epochs.
In Section \ref{sec:modeling}, we model the multi-band light curves
of iPTF~16asu using different models. We discuss our results in Section
\ref{sec:discussion} and draw some conclusions in Section \ref{sec:Con}.

\section{The Blackbody Fits for SEDs of iPTF~16asu}
\label{sec:SED}

W17 have used the blackbody model to fit the SEDs at the epochs when the
observations in at least three bands are available simultaneously
and the first two spectra of iPTF~16asu, reporting the evolution of temperature
and the radius of the photosphere of iPTF~16asu. However,
the derived blackbody luminosity of iPTF~16asu at the most epochs haven't been
presented. Here, we re-fit the SEDs at all epochs when at least three filter data are available.

To combine the $gri$ and photometry of $Swift$-UVOT, the $gri$ data at day $+3.51$ have
been interpolated to day $+3.47$.
We fit the unique UV--optical--IR SED (at day $+3.47$) by using the UV-absorbed blackbody model,
in which the optical--IR part of the SED is fitted by the equation
$F_{\nu,{\rm ph}}(\lambda > \lambda_{\rm cut}) =F_{\nu,{\rm bb}} = (2 {\pi} h{\nu}^3/c^2)
(e^{\frac{h{\nu}}{k_{\rm b}T_{\rm ph}}}-1)^{-1}\frac{R_{\rm ph}^2}{D_L^2}$,
while the UV part of SED is fitted by $F_{\nu,{\rm ph}}(\lambda \leq \lambda_{\rm cut}) =
\big(\frac{\lambda}{\lambda_{\rm cut}}\big)^{\beta}F_{\nu,{\rm bb}}$
(see, e.g., \citealt{Pra2017,Nich2017b,Yan2020}); here, $T_{\rm ph}$ is the temperature of the SN photosphere,
$R_{\rm ph}$ is the radius of the SN photosphere, $D_L$ is the luminosity distance of the SN,
$\lambda_{\rm cut}$ is the cutoff wavelength, $\beta$ is a power-law index \citep{Yan2020}.
\footnote{\cite{Pra2017} and \cite{Nich2017b} suppose that $\lambda_{\rm cut} =3000$ \AA, $\beta=1$.}
Although the SEDs at all other epochs do not have the UV photometry, we suggest that it is more
reasonable to suppose that they are UV-absorbed ones and invoke the UV-absorbed model to
fit them. The values of $\lambda_{\rm cut}$ and $\beta$ are taken to be the best-fitting
ones of the fit for the SED at day $+3.47$, because there are no UV data to constrain the
two parameters for the other epochs. For comparison, however, the standard blackbody model is also invoked to fit the same SEDs.
The Markov Chain Monte Carlo (MCMC) method using the \texttt{emcee} Python package \citep{Foreman-Mackey2013}
is used to obtain the best-fitting parameters and the 1-$\sigma$ range of the parameters.

\begin{figure}[tbph]
\begin{center}
\includegraphics[width=0.4\textwidth,angle=0]{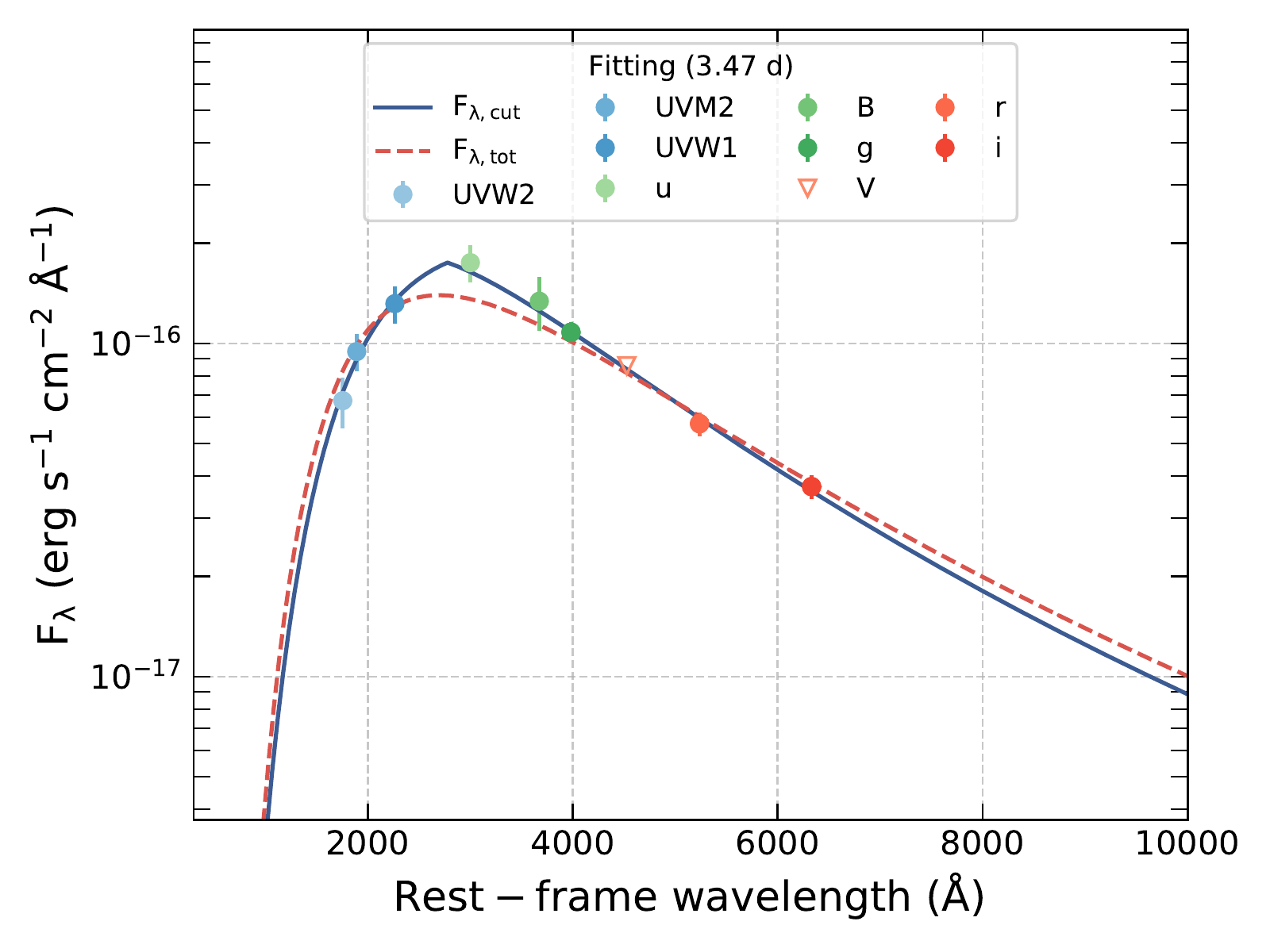}
\includegraphics[width=0.4\textwidth,angle=0]{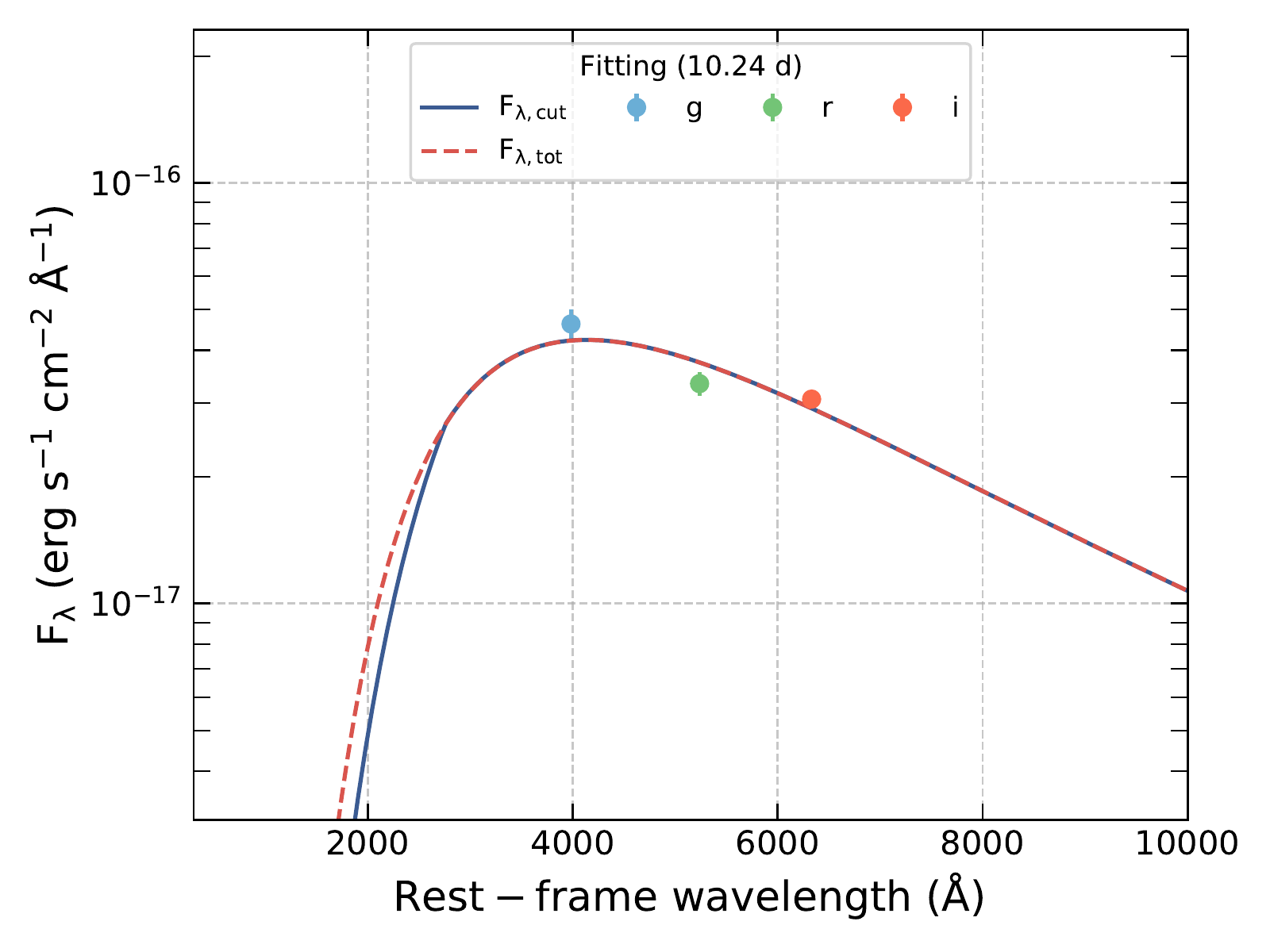}
\includegraphics[width=0.4\textwidth,angle=0]{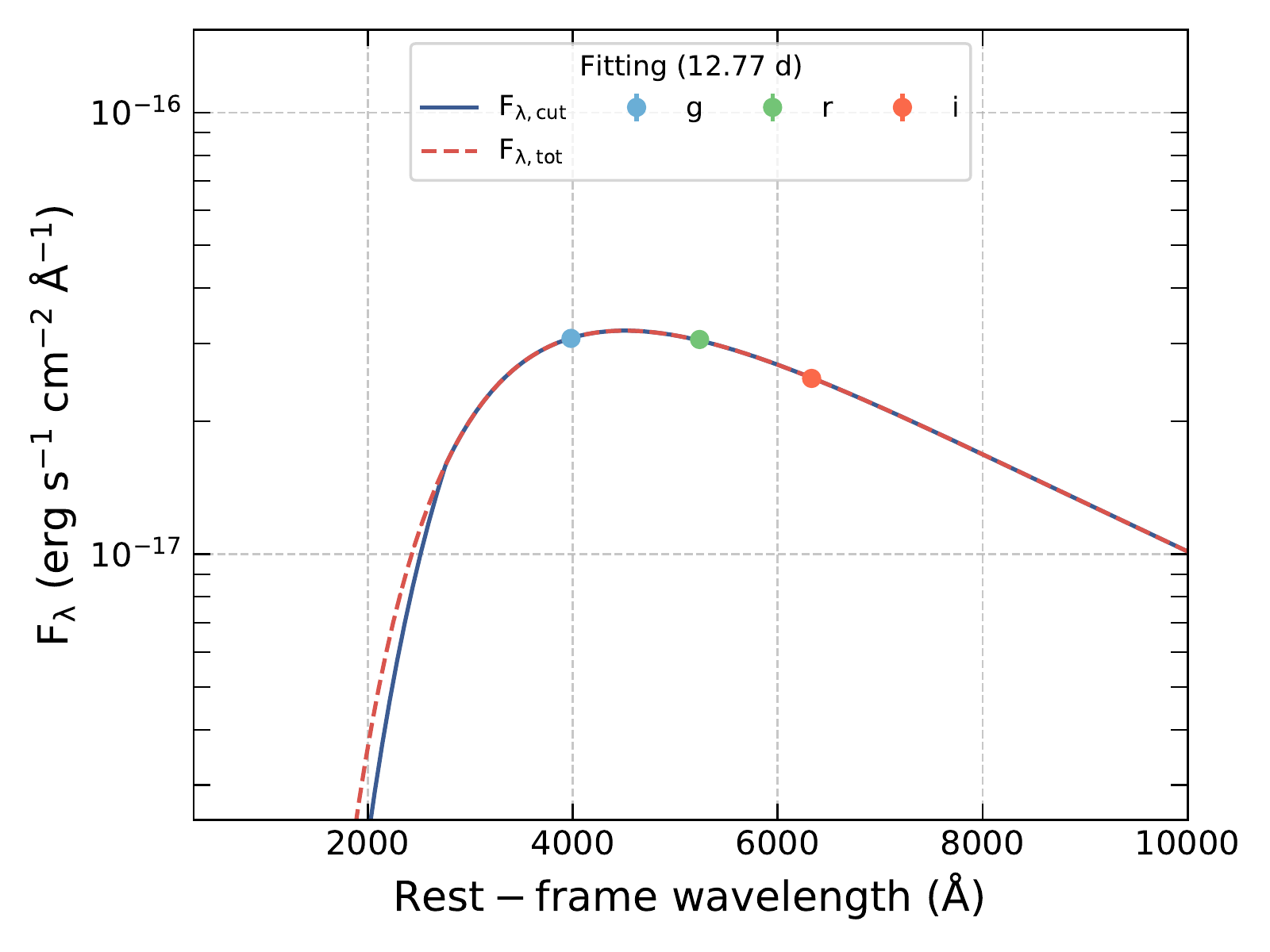}
\includegraphics[width=0.4\textwidth,angle=0]{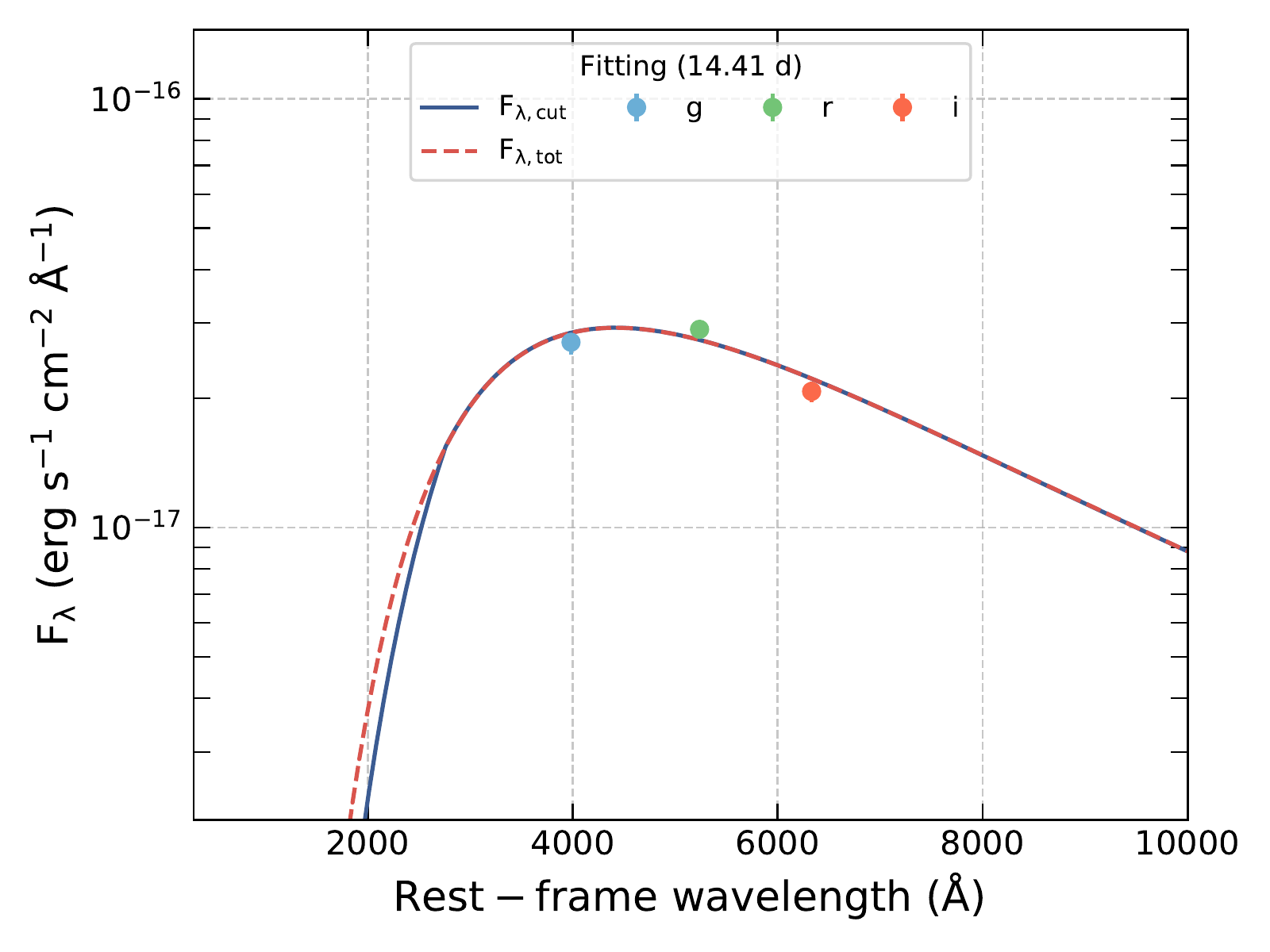}
\includegraphics[width=0.4\textwidth,angle=0]{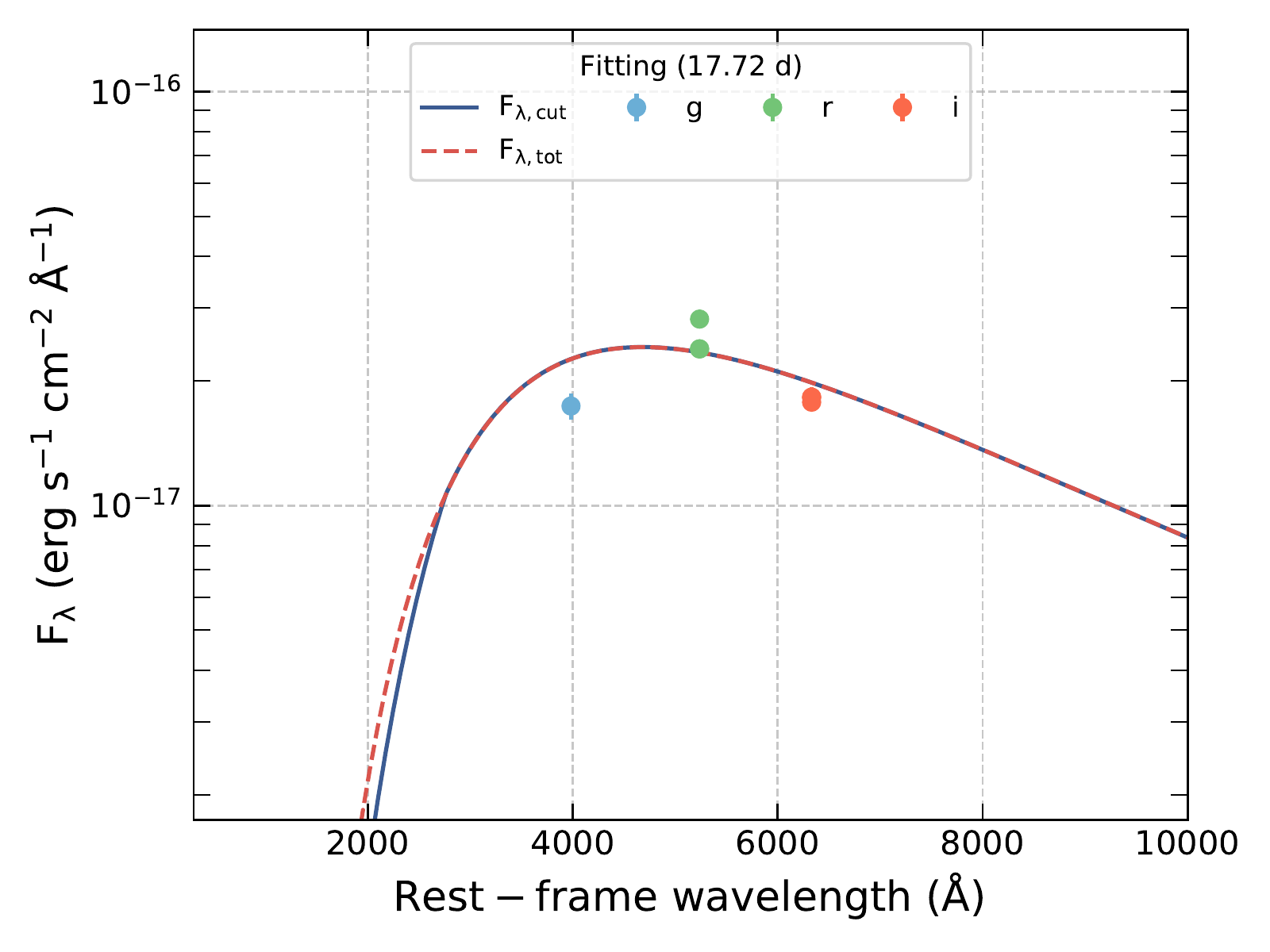}
\includegraphics[width=0.4\textwidth,angle=0]{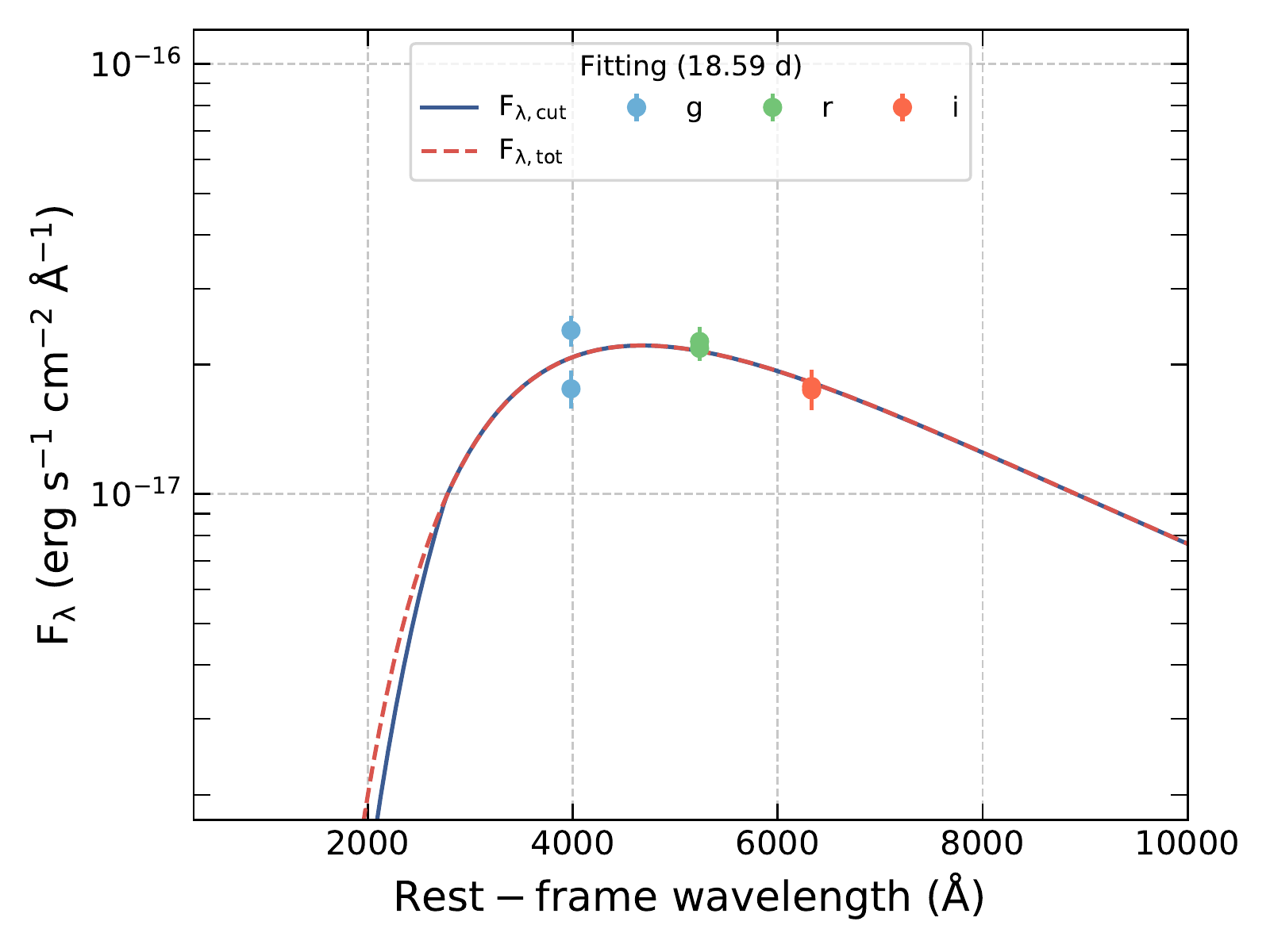}
\includegraphics[width=0.4\textwidth,angle=0]{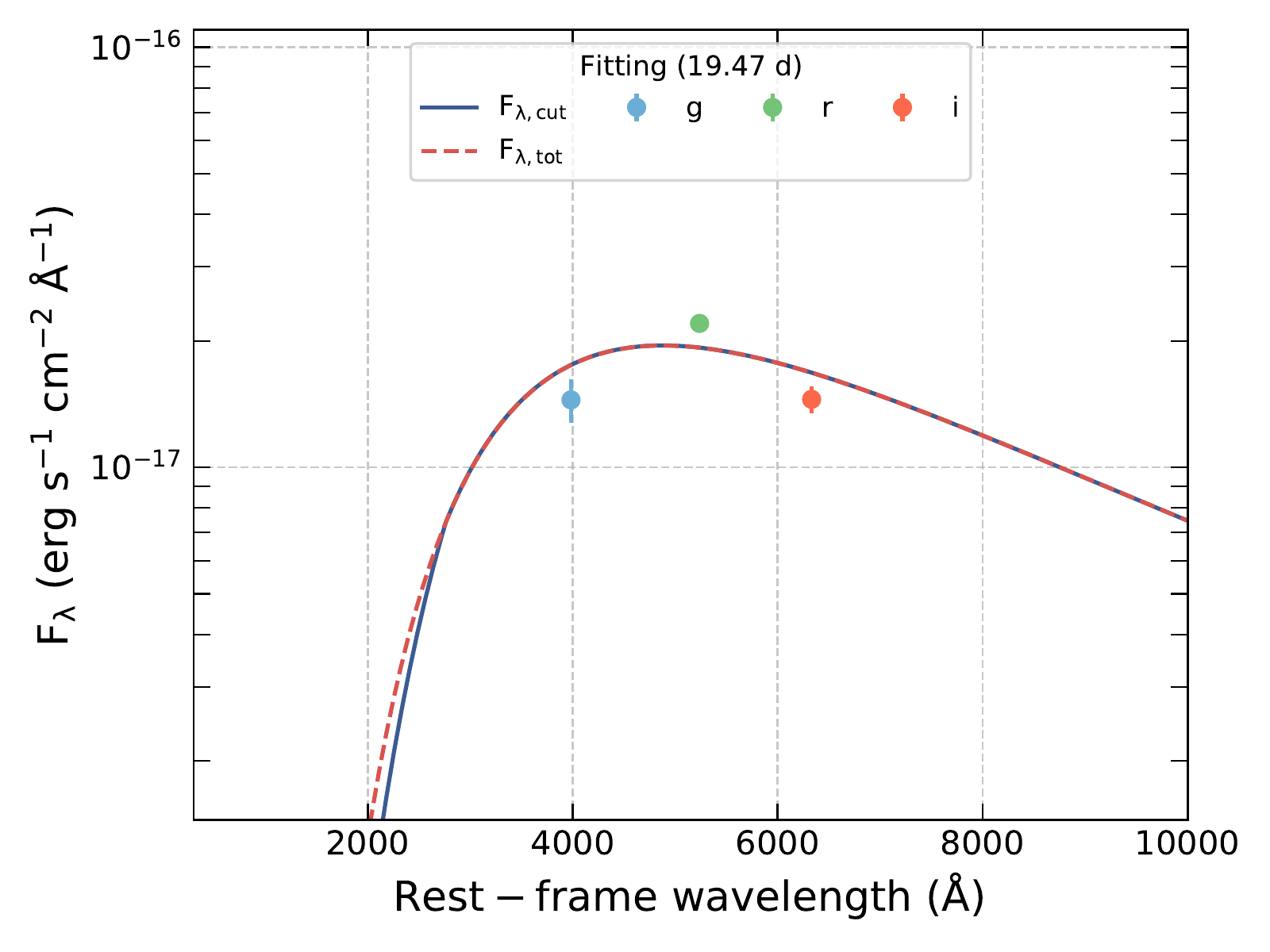}
\end{center}
\caption{The best fits of the SEDs of iPTF~16asu at all epochs.
The solid lines represent the theoretical SEDs produced by the UV-absorbed blackbody model.
For comparison, the fits using the standard blackbody model are plotted by the dashed lines.
The data are from Table 1 of W17, the triangle in the first panel represents the $V-$band upper limit.}
\label{fig:SED}
\end{figure}

We present the medians and 1-$\sigma$ bounds of the temperature and the radii
of the SN photosphere, as well as the bolometric luminosity derived by using
$L_{\rm ph} = \int_0^{\infty} F_{\nu,{\rm ph}}{\rm d}\nu$ at all epochs
in Table \ref{table:SED_L}, and plot all fits reproduced by
the UV-absorbed blackbody model (the solid lines) as well as the standard
blackbody model (the dashed lines) in Figure \ref{fig:SED}.

We find that the UV-absorbed blackbody model is better than the standard blackbody model,
because the former can fit all data at day $+3.47$ (see Figure \ref{fig:SED}) and,
as shown in Figure \ref{fig:SED}, the latter cannot fit the $u$-band photometry (see also Figure 6 of W17).
The value of the reduced $\chi^2$ ($\chi^2$/dof = 0.20, dof=degree of freedom) of the former
is also smaller than that ($\chi^2$/dof = 0.79) of the latter. The derived value of $\lambda_{\rm cut}$ is
$2757.81^{+429.64}_{-365.93}$ \AA, which is comparable to that of some type I SLSNe
($\sim$3000 \AA, e.g., \citealt{Chom2011}, \citealt{Pra2017}, \citealt{Nich2017a}, and \citealt{Nich2017b}).
The photosphere temperature at day $+3.47$ we derive
($12,004^{+743}_{-632}$ K, see Table \ref{table:SED_L}) is higher than that derived by W17 ($10,800\pm 250$ K);
in contrast, the photosphere radius at the same epoch we derived ($(2.18\pm 0.17)\times 10^{15}$ cm)
is smaller than that derived by W17 ($(2.6\pm 0.2)\times 10^{15}$ cm). Adopting
$L_{\rm ph} = \int_0^{\infty} F_{\nu,{\rm ph}}{\rm d}\nu$  (rather than
$L_{\rm ph} = 4 \pi \sigma T_{\rm ph}^4R_{\rm ph}^2$), we find that the bolometric luminosity
at day $+3.47$ is $6.20^{+0.19}_{-0.18}\times 10^{43}$ erg s$^{-1}$, which is slightly lower than
that derived by W17 ($(6.4\pm 1.6)\times 10^{43}$ erg s$^{-1}$),
but still about 2 times the peak ($(3.4\pm 0.3)\times 10^{43}$ erg s$^{-1}$)
of the pseudo-bolometric light curve constructed by W17.

\begin{figure}[tbph]
\begin{center}
\includegraphics[width=0.6\textwidth,angle=0]{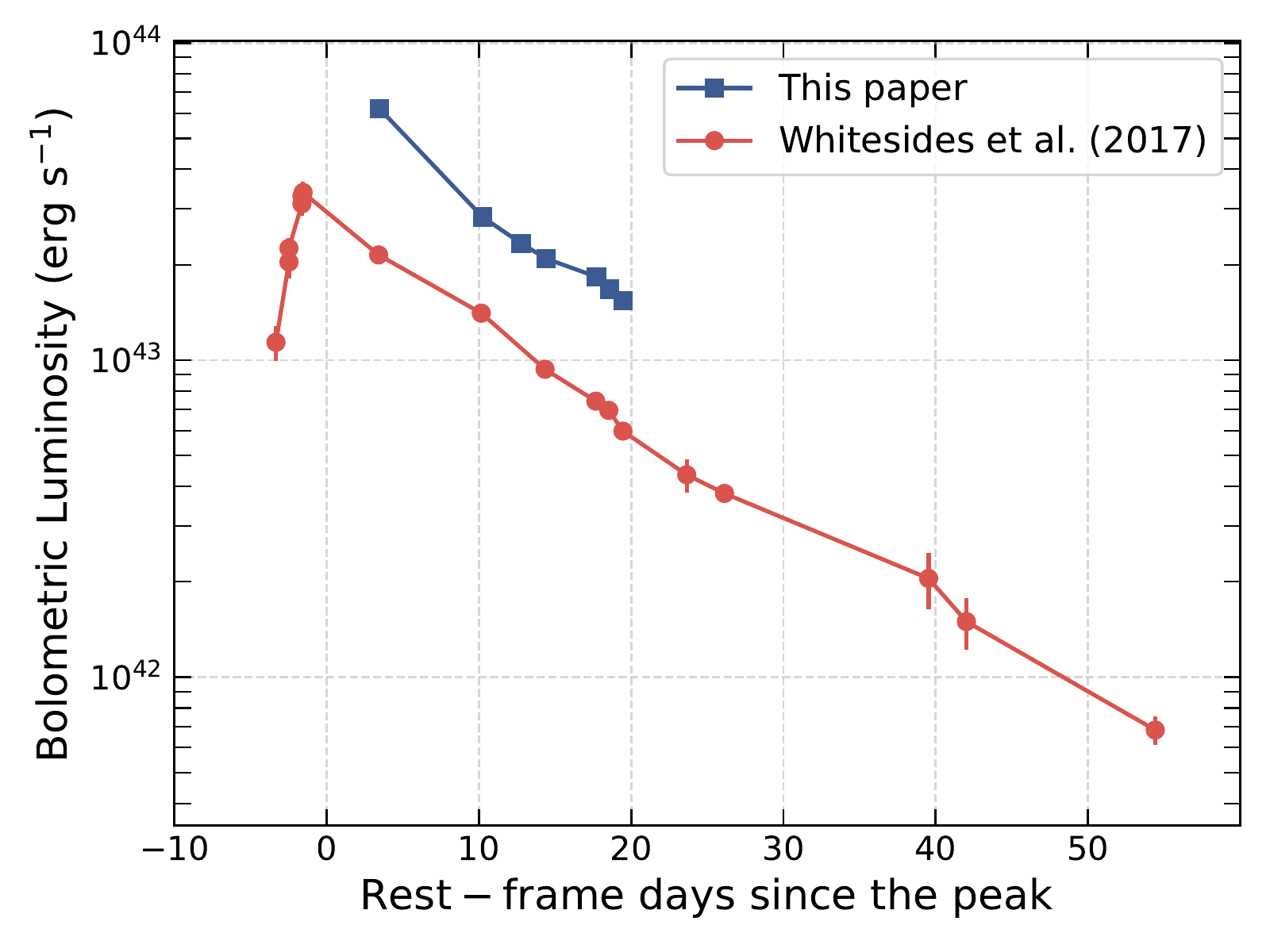}
\end{center}
\caption{The post-peak bolometric light curve derived from the SEDs
(the filled squares), the pseudo-bolometric light
curve constructed by W17 is represented by the filled circles.}
\label{fig:SED-L}
\end{figure}

We compare the post-peak bolometric light curve we derive from the SEDs and the
pseudo-bolometric light curve constructed by W17 by plotting them in Figure \ref{fig:SED-L}.
It can be found that the bolometric light curve we derive is significantly more luminous
than the pseudo-bolometric light curve constructed by W17. Especially, the extrapolated
peak luminosity of the bolometric luminosity of iPTF~16asu can exceed $\sim 10^{44}$ erg s$^{-1}$,
which is about 3 times the peak of the pseudo-bolometric light curve
($(3.4\pm 0.3)\times 10^{43}$ erg s$^{-1}$), and brighter than the threshold ($7\times 10^{43}$ erg s$^{-1}$,
\cite{Gal2012}) of SLSN. This indicates that iPTF~16asu might be a genuine SLSN.

\section{Modeling the Multi-band Light Curves of iPTF~16asu}
\label{sec:modeling}

The fact that the bolometric luminosity inferred from the UV-absorbed blackbody
fits of the SEDs is higher than the pseudo-bolometric light curve constructed by W17
indicates that the energy source powering the luminosity evolution must be
re-investigated.

We first use the $^{56}$Ni model to fit the multi-band light curves. The details
of the $^{56}$Ni model reproducing the bolometric light curves of SNe can be found in
\citet{Wang2015b} and references therein. \footnote{Note that the
factor $(1-e^{-\tau_{\gamma}(t)})$ in Equation (1) of \citet{Wang2015b} and other
literature (e.g., \citealt{Cha2012}, \citealt{Cha2013}, \citealt{Wang2015a},
\citealt{Nich2017b}) that presents the $\gamma$-ray trapping factor ought to be
inside, rather than outside, the integral.}
Throughout this paper, the optical opacity of the ejecta $\kappa$ is
taken to be 0.10 cm$^2$~g$^{-1}$.
To fit the multi-band light curves, the photosphere evolution
modules must be incorporated into the model, as done by \citet{Nich2017b} for the
fits of the multi-band light curves of SLSNe, see their Equations (8) and (9).
The equations determining SEDs are presented in Section \ref{sec:SED}.
The definitions, the units, and the priors of the parameters of
the $^{56}$Ni model are listed in Table \ref{tab:Nimag-parameters}.
\footnote{The values of $\lambda_{\rm cut}$ and $\beta$ are taken
to be those derived by the fit for the first SED, see Section \ref{sec:SED}.}
The MCMC method using the \texttt{emcee} Python package
\citep{Foreman-Mackey2013} is also used here.

\begin{figure}[tbph]
\begin{center}
\includegraphics[width=0.70\textwidth,angle=0]{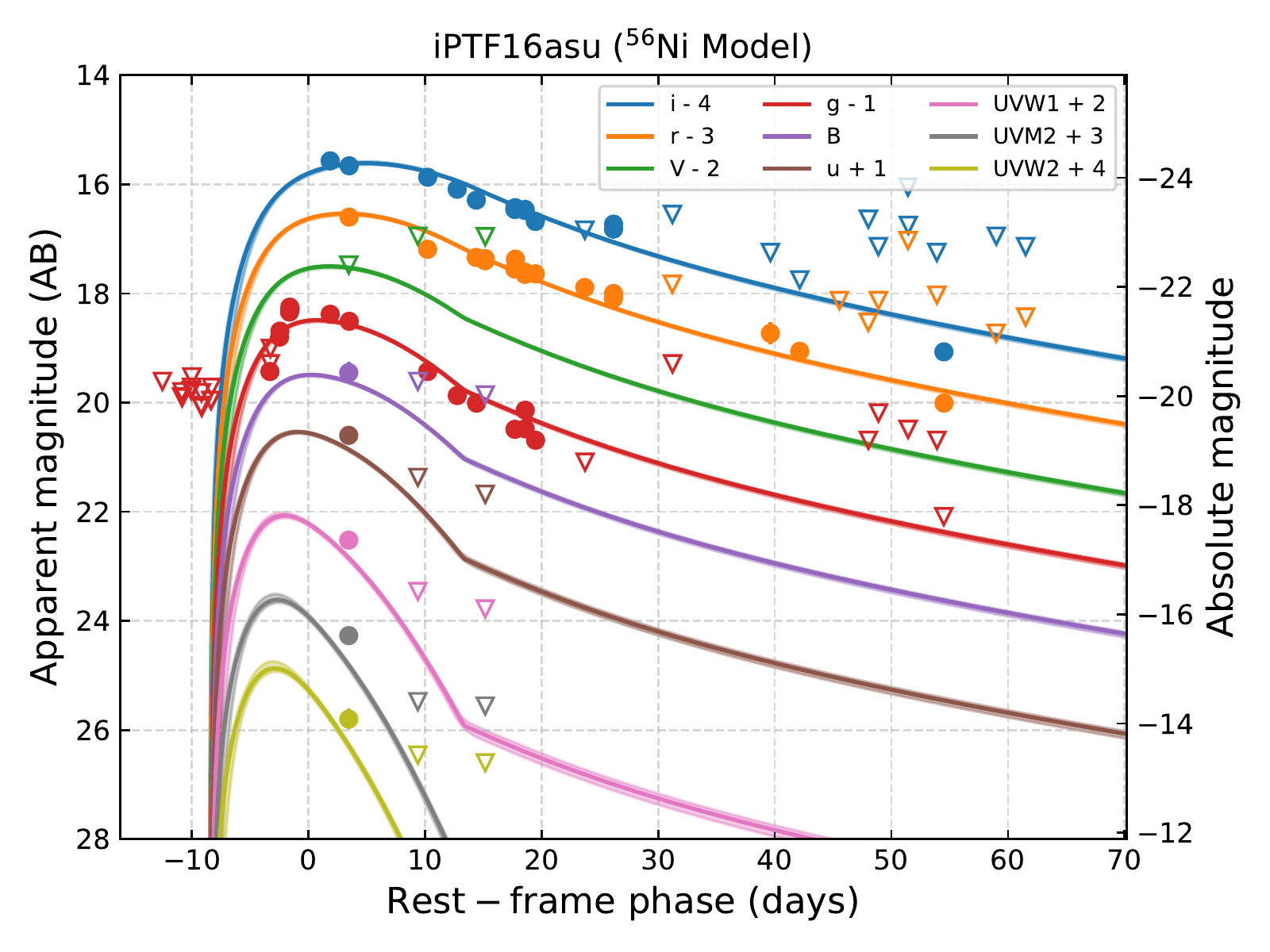}
\includegraphics[width=0.70\textwidth,angle=0]{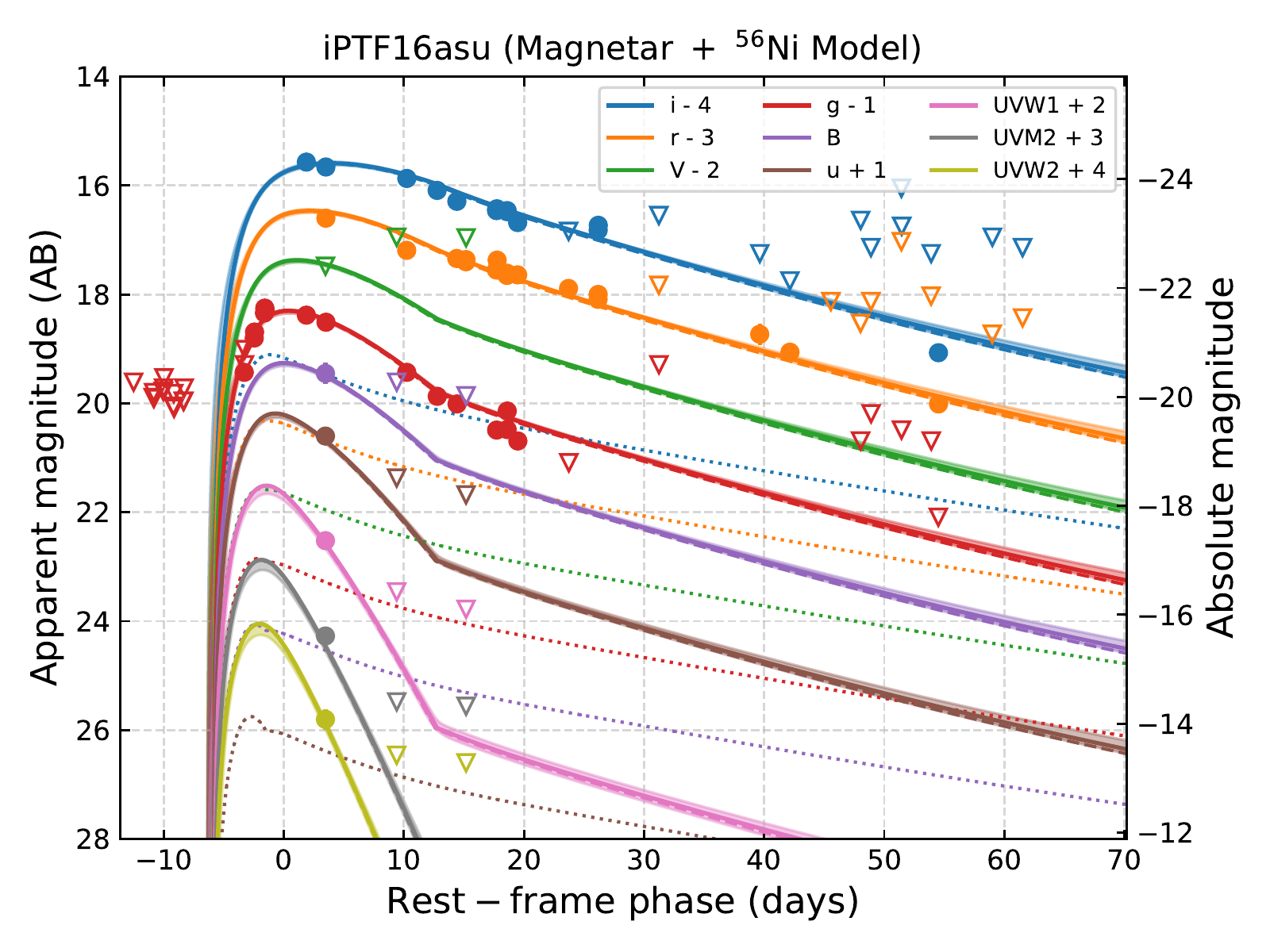}
\end{center}
\caption{The best fits (the solid curves) of the multi-band light curves
of iPTF~16asu using the $^{56}$Ni model (the top panel) and the magnetar plus $^{56}$Ni model
(the bottom panel), respectively. The shaded regions indicate 1-$\sigma$ bounds
of the parameters. The dotted lines and dashed lines in the bottom panel
are the light curves powered by the $^{56}$Ni and the magnetar, respectively.
The data are from Table 1 of W17, triangles represent upper limits.}
\label{fig:multibandfits}
\end{figure}

The $^{56}$Ni model cannot match the pre-peak $g-$band light curve,
see Figure \ref{fig:multibandfits}. Moreover, the derived $^{56}$Ni is
$1.98_{-0.03}^{+0.01}$~M$_\odot$, which is $\sim 3.6$ times that
derived by W17 ($0.55$~M$_\odot$), and significantly higher than
the ejecta mass ($0.30_{-0.07}^{+0.08}$~M$_\odot$, see Table \ref{tab:Nimag-parameters},
or Figure \ref{fig:corner_Ni}). This supports W17's conclusion that the $^{56}$Ni model
is disfavored.

The magnetar model is one of the most prevailing models used
to account for the SNe that cannot be explained by the $^{56}$Ni model.
For completeness, however, the contribution of $^{56}$Ni is also
included and the ratio of the $^{56}$Ni 
to the ejecta mass is supposed to be $\leq$ 0.2 \citep{Ume2008}.
The details of the magnetar plus $^{56}$Ni model we adopt can be found in
\citet{Wang2015b} and references therein.
The photosphere evolution modules are also from Equations (8) and (9) of \citet{Nich2017b}.
We suppose that the mass and radius of the magnetar are
respectively $1.4$~M$_\odot$ and 10 km and list the free parameters of the magnetar plus $^{56}$Ni
model, their definitions, as well as the priors in Table \ref{tab:Nimag-parameters}.

The fit produced by the magnetar plus $^{56}$Ni model is also shown
in Figure \ref{fig:multibandfits}.
The parameters and the corresponding corner plot are presented in
Table \ref{tab:Nimag-parameters} and Figure \ref{fig:corner_magni}, respectively.
The magnetar plus $^{56}$Ni model can better fit the pre-peak $g-$band light curve,
and the value of $\chi^2$/dof is 4.31, which is smaller than that (6.61) of the  
$^{56}$Ni model.

\section{Discussion}
\label{sec:discussion}

\subsection{The Analysis for the Parameters of the Magnetar plus $^{56}$ Ni model}

The ejecta mass derived by the magnetar model is $0.21_{-0.06}^{+0.08}$~M$_\odot$, which
is $\sim 2-3$ times that derived by W17. Although it is still smaller than those of
other SLSNe, we suggest that this value is not problematic since it is comparable with
those of ultra-stripped SNe \citep{Tauris2013,Tauris2015}.
\footnote{For comparison, \cite{De2018} show that the ejecta mass of iPTF~14gqr which
is a fast-evolving SN Ic is $\sim 0.2$~M$_\odot$; \cite{Yao2020} study a rapidly evolving
SN~Ib SN~2019dge, and find that its ejecta mass is $\sim\,0.3$\,M$_\odot$; \cite{Prit2021}
study SN~2018gep which is another luminous ($M_{\rm r,peak}=-19.49\pm0.23$ mag) fast-evolving
($t_{\rm rise,V}\lesssim 6.2\pm 0.8$ days) SN Ic-BL with spectra resembling those of iPTF~16asu,
and show that its ejecta mass might be $\sim 0.26$~M$_\odot$ for the magnetar+$^{56}$Ni model
or $\sim 0.49$~M$_\odot$ for the ejecta--circum-stellar medium (CSM) interaction plus $^{56}$Ni
model.} Furthermore, the ejecta can be larger if we suppose that the value of $\kappa$ was
0.05$-$0.07 cm$^2$ g$^{-1}$, instead of the value we adopt.
Therefore, we suggest that iPTF~16asu might be an ultra-stripped SLSN Ic whose
luminosity was boosted by a nascent magnetar.

The derived $``$floor temperature$"$ ($6410.26^{+114.72}_{-108.19}$ K)
is consistent with that derived by W17 ($\sim 6000$ K,
see their Figure 7) from the fits for the SEDs of iPTF~16asu
as well as many other SLSNe I (see Table 3 of \citealt{Nich2017b}).
On the other hand, the temperature flattened to the $``$floor temperature$"$
after $\sim$ 18 days after the peak, which
is significantly smaller than those of other SLSNe ($\sim$ 50 days, see
\citealt{Nich2017b}). The reason is that the ejecta mass is very small
and the photosphere shrunk more quickly than those of other SLSNe.

The early-time photosphere velocity inferred is $(2.91\pm 0.16)\times 10^9$ cm s$^{-1}$,
slightly lower than the value ($(3.45\pm 0.54)\times 10^9$ cm s$^{-1}$) given by W17.
\footnote{In principle, the velocity of the photosphere is lower than the
velocity inferred from absorption lines of the spectra, because the latter are formed
in the SN atmospheres which are above the photospheres and have larger expansion
velocity, see the discussion in \cite{Wang2022}.}
The derived rising time is $5.97_{-0.42}^{+0.34}$ days, slightly
larger than that derived using the second-order polynomial fit
($3.97\pm 0.19$ days, W17).

The small ejecta mass and the high velocity result in a short diffusion time and
rise time, then the needed input power is lower than those of ``normal$"$ SLSNe
(according to the ``Arnett law$"$, \citealt{Arn1982}). Therefore, the best-fitting
$P_0$ (9.79 ms) is significantly larger than those of most SLSNe ($\sim 1-5$ ms).

\subsection{The Theoretical Bolometric Light Curve, the Temperature Evolution, and the Radius Evolution of iPTF~16asu}

\begin{figure}[tbph]
\begin{center}
\includegraphics[width=0.48\textwidth,angle=0]{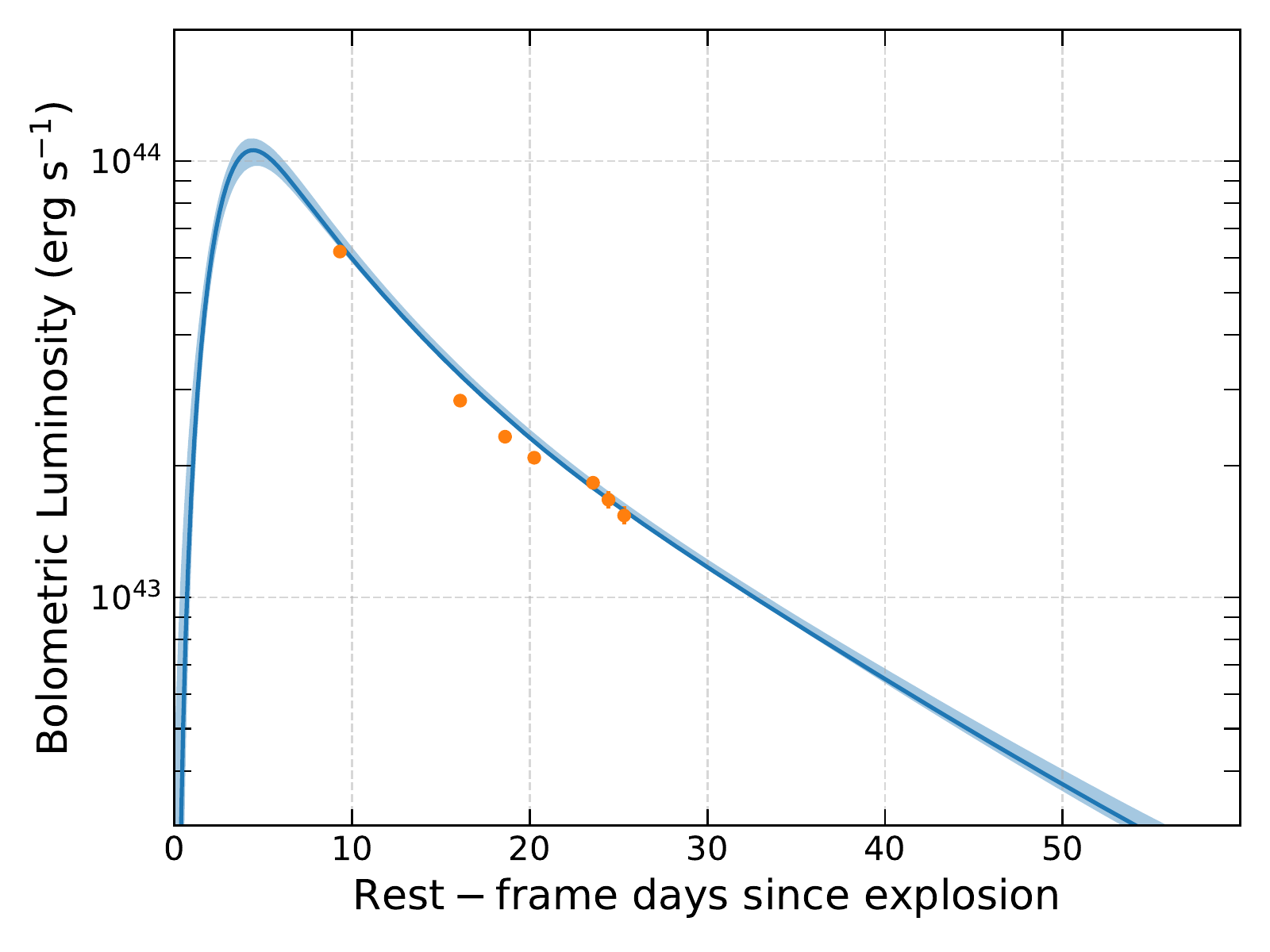}
\includegraphics[width=0.48\textwidth,angle=0]{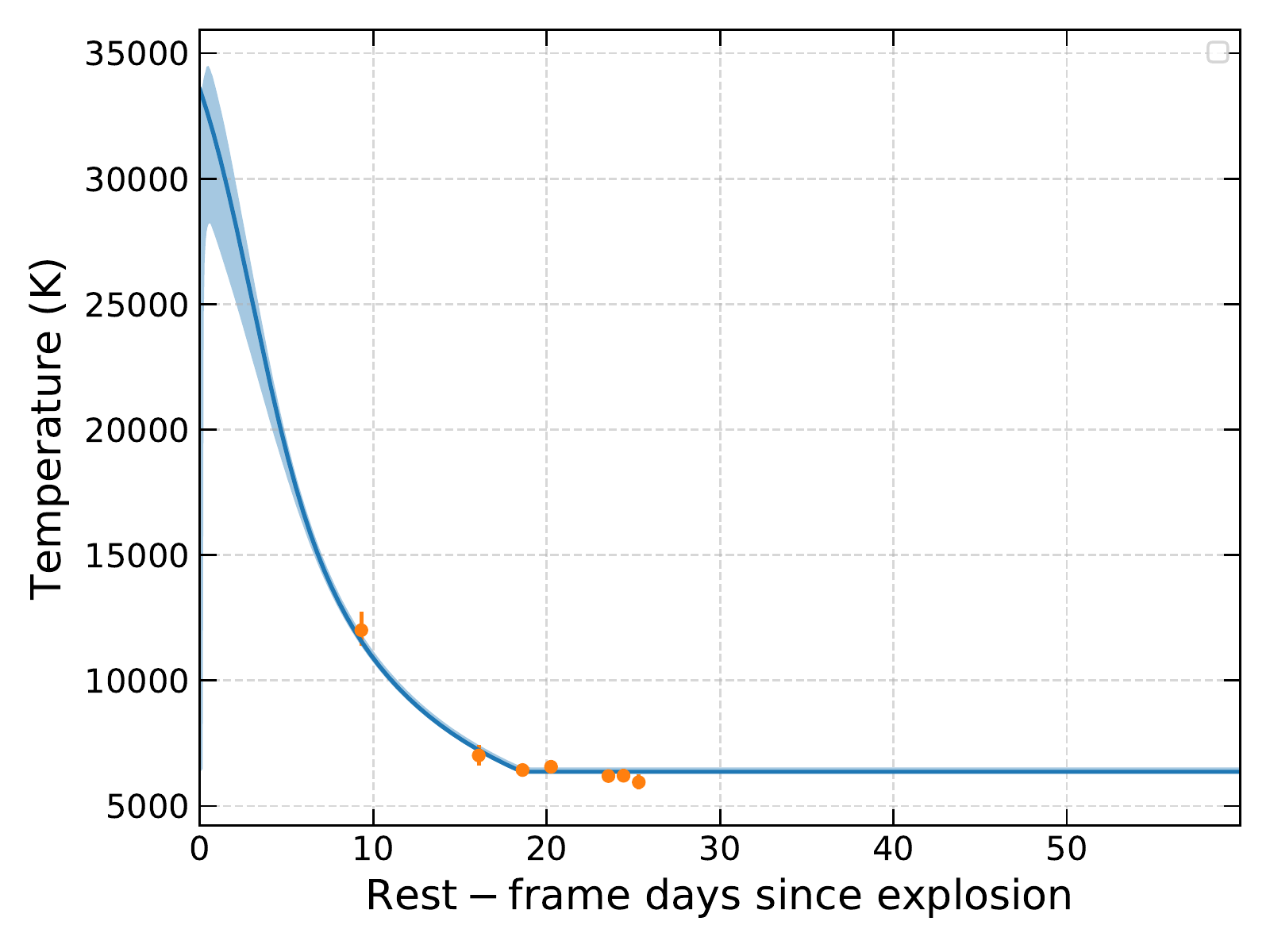}
\includegraphics[width=0.48\textwidth,angle=0]{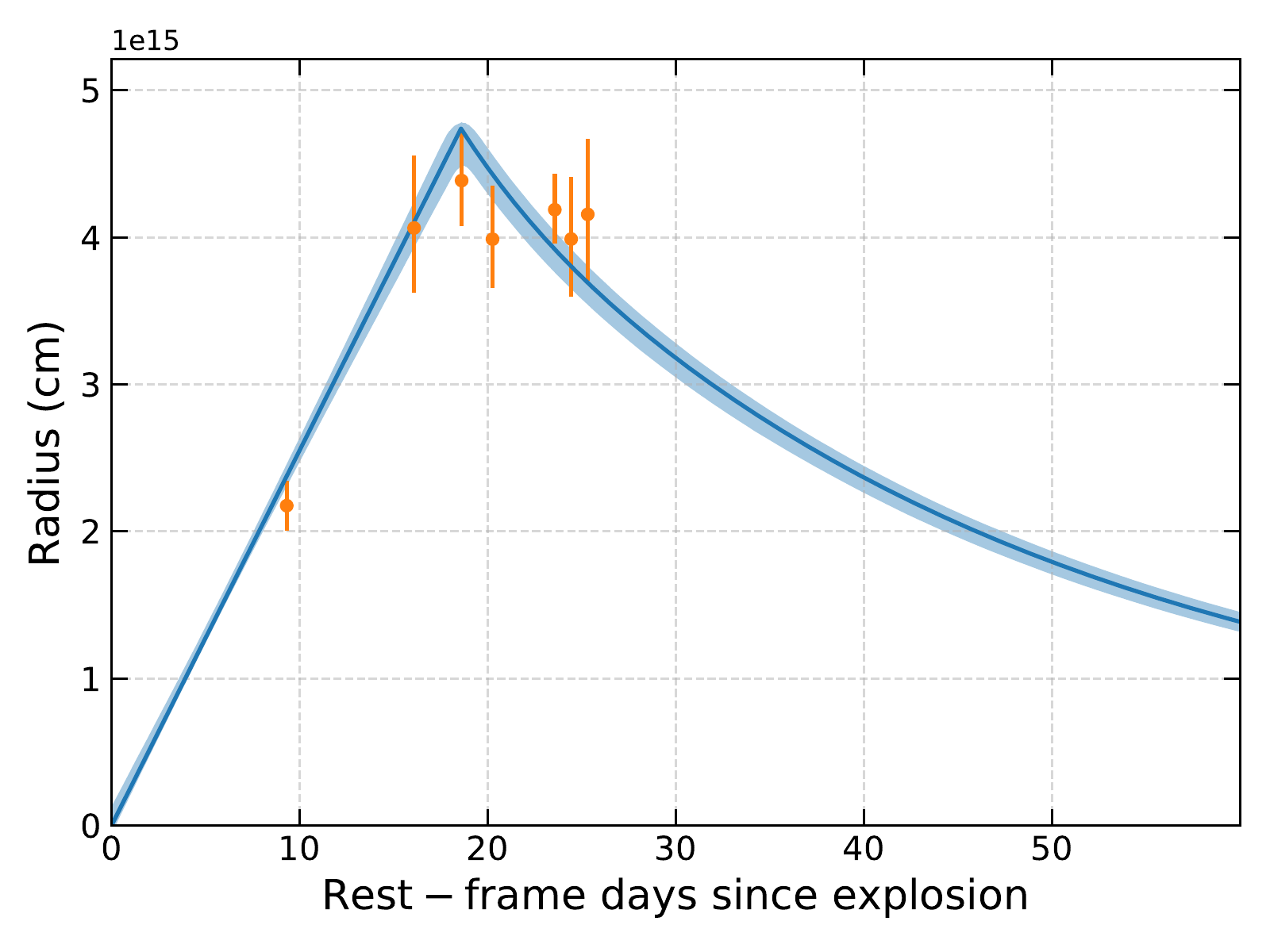}
\end{center}
\caption{The bolometric light curve, the temperature evolution, and the radius evolution reproduced by
the magnetar plus $^{56}$Ni model. The shaded regions indicate 1-$\sigma$ errors of the parameters.
Also plotted are the corresponding values (see Table \ref{table:SED_L}) derived from the observations.}
\label{fig:bolo}
\end{figure}

In section \ref{sec:SED}, we get the post-peak bolometric light curve of iPTF~16asu by
fitting the SEDs at seven epochs and using $L_{\rm ph} = \int_0^{\infty} F_{\nu,{\rm ph}}{\rm d}\nu$
to calculate the luminosity at the epochs. To obtain the full bolometric light curve, we use
the theoretical multi-band light curves produced by the best-fitting parameters to construct its
bolometric light curve by integrating the theoretical SEDs at all epochs, see the top-left panel of
Figure \ref{fig:bolo}. We also plot the temperature evolution and the radius evolution
reproduced by the magnetar plus $^{56}$Ni model in Figure \ref{fig:bolo}
(see the top-right and the bottom panels, respectively).
For comparison, bolometric luminosity, the temperature evolution and the radius evolution
derived by the photometry (see Table \ref{table:SED_L}) are also plotted in the same figure.

The peak luminosity of the bolometric light curve we construct is $\sim 1.06\times 10^{44}$
erg s$^{-1}$, which is $\sim 3$ times the peak luminosity derived by W17 ($(3.4\pm 0.3)\times10^{43}$ erg s$^{-1}$),
indicating that iPTF~16asu is NOT a luminous SN between SLSNe and normal SNe, but a genuine SLSN
even if the most stringent threshold ($>7\times 10^{43}$ erg s$^{-1}$, \citealt{Gal2012}) was adopted.
Using trapezoidal integration over period from the explosion to 100 days after explosion,
we calculate the total radiated energy of $\sim 1.28\times 10^{50}$ erg, which is
also about 3$-$4 times that calculated by W17 ($4.0\pm 0.6\times 10^{49}$ erg).

As pointed out by \cite{Inse2013}, using $L_{\rm ph} = 4 \pi \sigma T_{\rm ph}^4R_{\rm ph}^2$
would overestimate the bolometric luminosity.
However, our method deriving the theoretical bolometric luminosity at all epochs
(including the epochs without any observations) can avoid overestimating the
bolometric luminosity, since we adopt the UV-absorbed blackbody model
\footnote{Although only one SED has UV photometry, our UV-absorbed
fit for the multi-band light curves of iPTF~16asu was applied for all epochs.}
and calculate the luminosity via the equation
$L_{\rm ph} = \int_0^{\infty} F_{\nu,{\rm ph}}{\rm d}\nu$,
rather than $L_{\rm ph} = 4 \pi \sigma T_{\rm ph}^4R_{\rm ph}^2$.

\section{Conclusions}
\label{sec:Con}

iPTF~16asu has so far been classified as a luminous rapidly evolving SN Ic-BL whose
multi-band light curves resembles those of luminous REOTs. In this paper, we re-analyze the SEDs of iPTF~16asu and re-construct
its post-peak bolometric light curve. We find that the bolometric luminosity at $+3.47$ days
after the peak is $6.20^{+0.19}_{-0.18}\times 10^{43}$ erg s$^{-1}$. Although this value is
slightly lower than the value ($(6.4\pm1.6)\times 10^{43}$ erg s$^{-1}$) derived by W17
, it still significantly brighter than the peak of pseudo-bolometric light curve constructed by W17
($(3.4\pm0.3)\times10^{43}$ erg s$^{-1}$).

Extrapolating the post-peak bolometric light curve we derive would result in a peak
luminosity of $\sim 10^{44}$ erg s$^{-1}$, which exceeds the threshold of SLSNe
($>7\times10^{43}$ erg s$^{-1}$, \citealt{Gal2012}). This fact indicates that the luminosity of the
bolometric light curve of iPTF~16asu might be about 3 times that of the pseudo-bolometric 
light curve constructed by W17, and iPTF~16asu might be a SLSN.

We use the $^{56}$Ni model and the magnetar plus $^{56}$Ni model to
fit the multi-band light curves, and construct the theoretical bolometric light curves by
integrating the theoretical SEDs constructed by the theoretical multi-band light curves. 
Our modeling disfavors the $^{56}$Ni model and favors the magnetar plus $^{56}$Ni model. 
Although the ejecta mass we derived is $0.21_{-0.06}^{+0.08}$~M$_\odot$ which is very low,
we suggest that this value is not problematic, since it is comparable with the ejecta masses
of ultra-stripped SNe, and a smaller value of $\kappa$ would give a larger ejecta
mass. It is reasonable to expect a magnetar boosted the luminosity of iPTF~16asu and made it
an ultra-stripped/rapidly evolving SLSN Ic-BL. 

The bolometric light curve derived by the theoretical SEDs extracted from the theoretical
multi-band light curves reproduced by the best-fitting parameters of the magnetar plus
$^{56}$Ni model shows that its peak is $\sim 1.06\times 10^{44}$ erg s$^{-1}$,
indicating that iPTF~16asu is indeed a rapidly evolving
SLSN Ic-BL, since the peak luminosity is higher than the threshold of SLSNe. This conclusion
is robust, since we use the UV-absorbed blackbody model to obtain the theoretical SEDs at all
epochs, and avoid overestimating the bolometric luminosity.

Our work further highlights the importance and robustness of the method directly fitting
the multi-band light curves of SNe and other optical transients. Especially, it is very
difficult to construct the early-time (pseudo-)bolometric light curves for a fraction of
SNe and other optical transients discovered by various wide-field sky-survey telescopes,
since only one or two band observations for them are available at the early epochs.
For the SNe and other optical transients having sparse early-time data,
modeling the multi-band light curves can yield more reliable results.

\acknowledgments

We thank the anonymous referee for helpful comments and
suggestions that have allowed us to improve this manuscript.
This work is supported by National Natural Science Foundation of China
(grant 11963001).

\clearpage

\begin{table*}
\centering
\tabletypesize{\scriptsize}
\caption{\label{table:SED_L}The medians and 1-$\sigma$ bounds of the parameters of the UV-absorbed Blackbody model for SEDs of iPTF16asu.}
\begin{tabular}{c c c c c c c}
\hline\hline
\colhead{Phase} & \colhead{$T_{\rm ph}$} & \colhead{$R_{\rm ph}$} & \colhead{$\lambda_{\rm cut}$} & \colhead{$\beta$}  & \colhead{$L_{\rm bolo}$}      &  \colhead{$\chi^{\rm 2}$/dof}  \\
(days)          & ($\rm 10^{3}$ K)       & ($\rm 10^{15}$ cm)     & ($\rm \mathring{A}$)         &         --                  & ($\rm 10^{43}\ erg\ s^{-1}$ ) &    --   \\

\hline
3.47 & $12.01^{+0.73}_{-0.63}$ & $2.18^{+0.17}_{-0.17}$ & $2758.56^{+432.87}_{-366.65}$ & $1.5^{+0.79}_{-0.62}$ & $6.20^{+0.19}_{-0.18}$ & 0.2\\
10.24 & $7.01^{+0.43}_{-0.39}$ & $4.06^{+0.49}_{-0.44}$ & --   & -- & $2.82^{+0.10}_{-0.10}$  & 5.89  \\
12.77 & $6.43^{+0.21}_{-0.2}$ & $4.39^{+0.34}_{-0.31}$ & --  & -- & $2.33^{+0.08}_{-0.07}$  & 0.024 \\
14.41 & $6.56^{+0.27}_{-0.25}$ & $3.99^{+0.36}_{-0.33}$ & -- & -- & $2.09^{+0.07}_{-0.06}$  & 4.54  \\
17.72 & $6.19^{+0.16}_{-0.16}$ & $4.19^{+0.24}_{-0.23}$ & -- & -- & $1.83^{+0.03}_{-0.03}$  & 13.26 \\
18.59 & $6.21^{+0.28}_{-0.26}$ & $3.99^{+0.42}_{-0.39}$ & -- & -- & $1.67^{+0.08}_{-0.07}$  & 1.63  \\
19.47 & $5.95^{+0.31}_{-0.28}$ & $4.16^{+0.51}_{-0.45}$ & -- & -- & $1.54^{+0.08}_{-0.07}$  & 14.96 \\

\hline

\noalign{\smallskip}
\end{tabular}
\end{table*}

\clearpage

\begin{table*}
\tabletypesize{\scriptsize}
\caption{The Definitions, the units, the prior, the medians, 1-$\sigma$ bounds, and the best-fitting values for the parameters of the $^{56}$Ni and the magnetar plus $^{56}$Ni models. The values of $\chi^{\rm 2}$/dof (reduced $\chi^{\rm 2}$) are also presented.}
\label{tab:Nimag-parameters}
\hspace{-30pt}
\begin{tabular}{c c c c c c c c}
\hline\hline
                                & \colhead{Definition}                                 &  \colhead{Unit}      &    \colhead{Prior}   & \colhead{Best fit} & \colhead{Median}\\
\hline

{\bf $^{56}$Ni}\\

\hline
$M_{\rm ej}$                    & the ejecta mass                                      &   M$_\odot$          &    $[0.1, 50]$      &   0.35 & $0.3^{+0.08}_{-0.07}$ \\
$v$                             & the ejecta velocity                                  &   $10^9$ cm s$^{-1}$ &    $[1.0, 5.0]$     &   2.48 & $2.53^{+0.12}_{-0.12}$ \\
$M_{\rm Ni}$                    & the $^{56}$Ni mass                                   &   M$_\odot$          &    $[0.0, 2.0]$     &  1.99 & $1.98^{+0.01}_{-0.03}$ \\
$\log \kappa_{\rm \gamma, Ni}$  & gamma-ray opacity of $^{56}$Ni-cascade-decay photons &   cm$^2$g$^{-1}$     &    $[-1.57, 4] $    & -0.78 & $-0.7^{+0.13}_{-0.11}$ \\
$T_{\rm f}$                     & the temperature floor of the photosphere             &   K                  &    $[3000, 10^4] $  &  6398.78 & $6397.14^{+104.1}_{-105.89}$ \\
$t_{\rm shift}$                 & the explosion time relative to the first data        &   days               &    $[-20, 0]$       &  -8.35 & $-8.1^{+0.4}_{-0.42}$ \\
$A_{\text{host,V}}$             & Extinction in the host galaxy                        &   mag                &    $[0, 1]$         &  0.001 & $0.0037^{+0.01}_{-0.0}$ \\
$\chi^{\rm 2}$/dof              &                                                      &                      &                     &  6.6 & 6.64 \\

\hline\hline

{\bf Magnetar + $^{56}$Ni}\\
\hline
$M_{\rm ej}$                    & the ejecta mass                                      &   M$_\odot$          &    $[0.1, 50]$      & 0.19 & $0.21^{+0.08}_{-0.06}$ \\
$v$                             & the ejecta velocity                                  &   $10^9$ cm s$^{-1}$ &    $[1.0, 5.0]$     & 2.95 & $2.91^{+0.16}_{-0.16}$ \\
$M_{\rm Ni}$                    & the $^{56}$Ni mass                                   &   M$_\odot$          &    [0.0, 0.2\,$M_{\rm ej}$]     & 0.029 & $0.018^{+0.02}_{-0.01}$ \\
$P_0$                           & the initial period of the magnetar                   &   ms                 &    $[0.8, 50]$       & 9.81     & $9.7^{+0.27}_{-0.37}$ \\
$B$                             & the magnetic field strength of the magnetar          &   $10^{14}$ G        &    $[0.1, 100]$      & 9.57 & $9.36^{+0.22}_{-0.22}$ \\
$\log \kappa_{\rm \gamma, Ni}$  & gamma-ray opacity of $^{56}$Ni-cascade-decay photons &   cm$^2$g$^{-1}$     &    $[-1.57, 4] $     & 0.41 & $0.34^{+0.18}_{-0.18}$ \\
$\log \kappa_{\rm \gamma, mag}$ & gamma-ray opacity of magnetar photons                &   cm$^2$g$^{-1}$     &    $[-2, 4] $        & 0.28 & $0.82^{+2.16}_{-1.52}$ \\
$T_{\rm f}$                     & the temperature floor of the photosphere             &   K                  &    $[3000, 10^4] $   & 6362.82 & $6408.44^{+114.73}_{-108.44}$ \\
$t_{\rm shift}$                 & the explosion time relative to the first data        &   days               &    $[-20, 0]$        & -5.84 & $-5.97^{+0.34}_{-0.42}$ \\
$A_{\text{host,V}}$             & Extinction in the host galaxy                        &   mag                &    $[0, 1]$          & 0.0071 & $0.021^{+0.03}_{-0.02}$ \\
$\chi^{\rm 2}$/dof              &                                                      &                      &                      & 4.3 & 4.36 \\
\hline\hline

\noalign{\smallskip}
\end{tabular}
\end{table*}


\clearpage

\clearpage

\clearpage

\appendix
\setcounter{table}{0}
\setcounter{figure}{0}
\setcounter{equation}{0}
\renewcommand{\thetable}{A\arabic{table}}
\renewcommand{\thefigure}{A\arabic{figure}}
\renewcommand\theequation{A.\arabic{equation}}

\begin{figure}[tbph]
\begin{center}
\includegraphics[width=0.63\textwidth,angle=0]{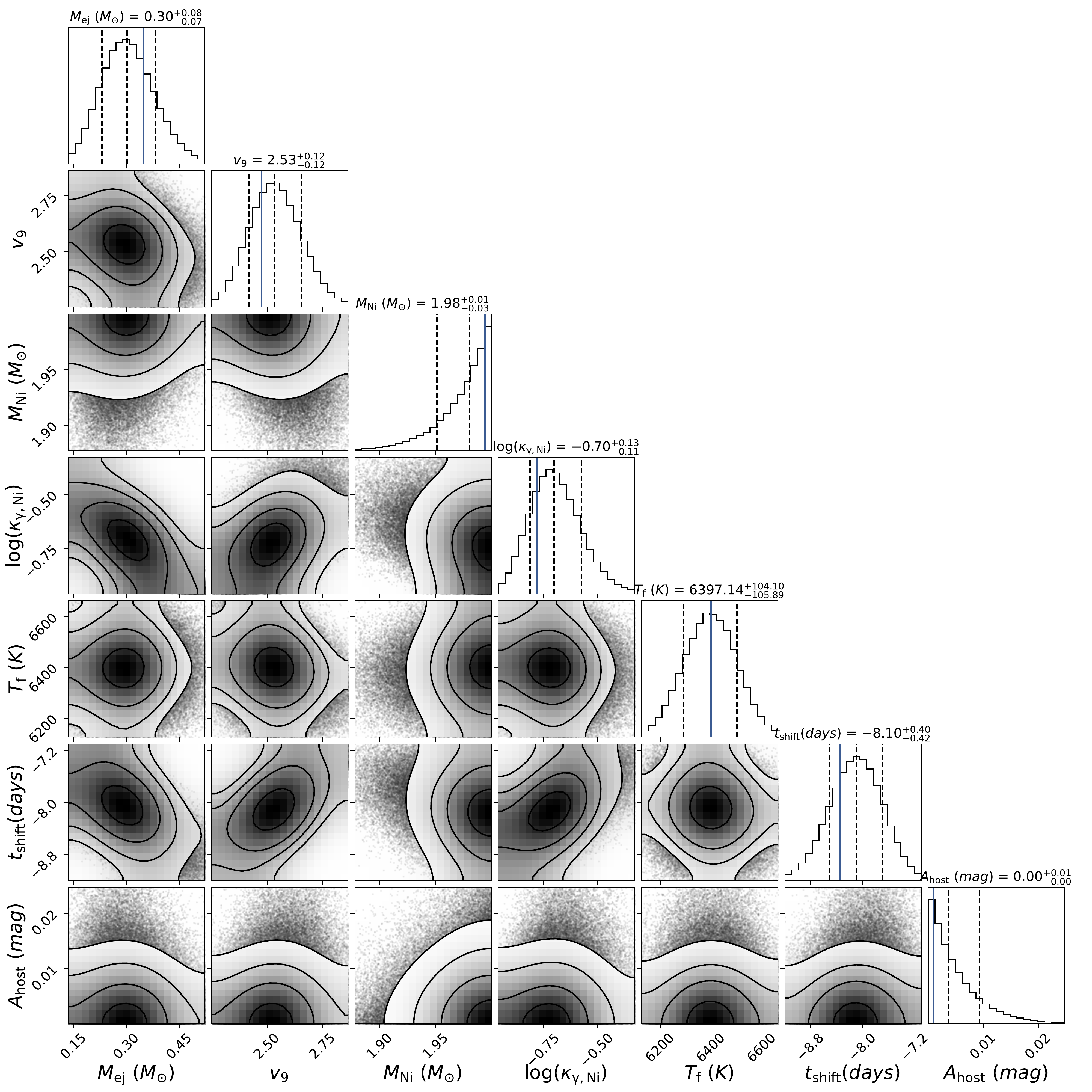}
\end{center}
\caption{The corner plot of the $^{56}$Ni model. The solid vertical lines represent the best-fit parameters, while the dashed vertical lines
represent the medians and the 1-$\sigma$ bounds of the parameters.}
\label{fig:corner_Ni}
\end{figure}

\begin{figure}[tbph]
\begin{center}
\includegraphics[width=0.9\textwidth,angle=0]{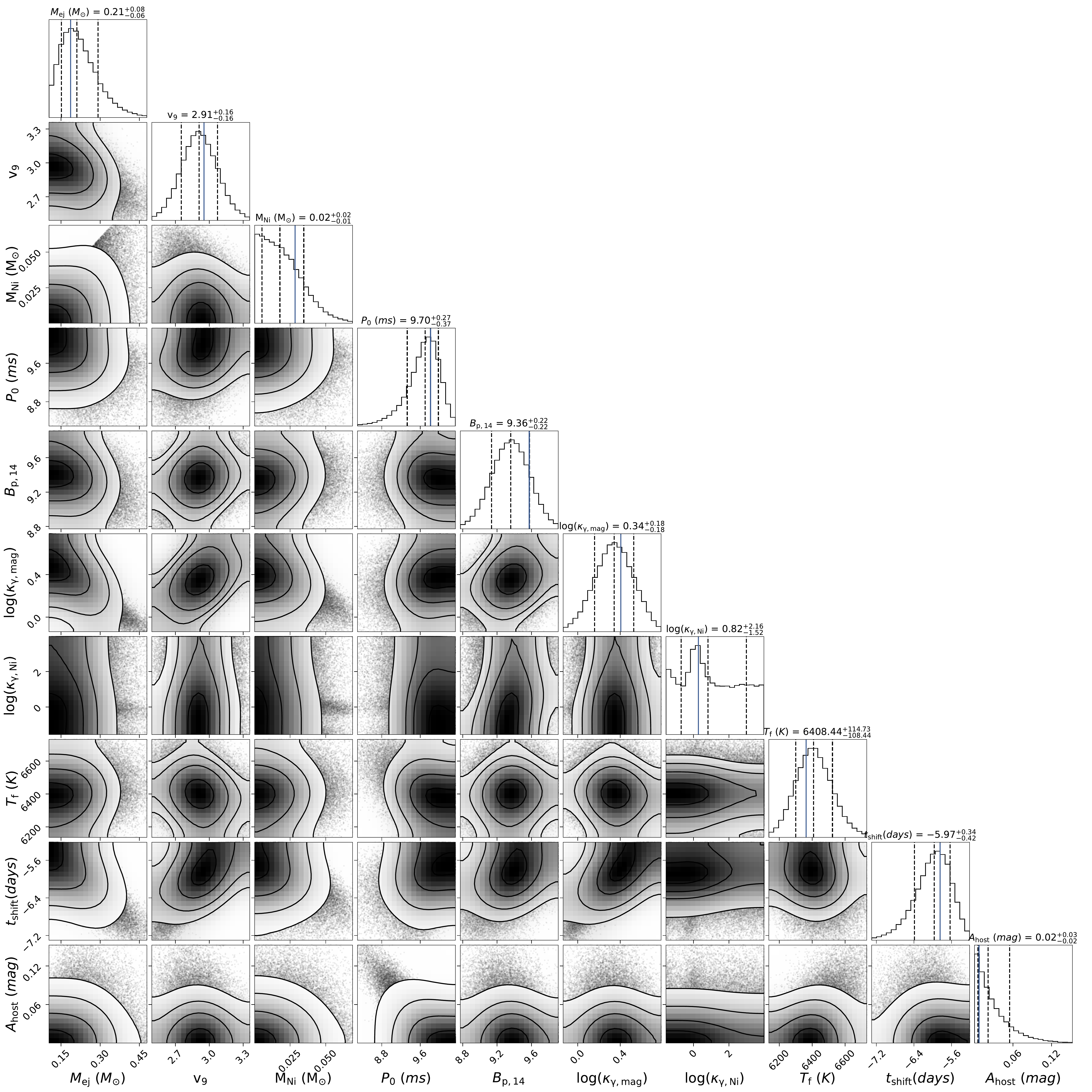}
\end{center}
\caption{The corner plot of the magnetar plus $^{56}$Ni model. The solid vertical lines represent the best-fit parameters, while the dashed vertical lines
represent the medians and the 1-$\sigma$ bounds of the parameters.}
\label{fig:corner_magni}
\end{figure}

Figures \ref{fig:corner_Ni} and \ref{fig:corner_magni}
present the corner plots of the $^{56}$Ni model
and the magnetar plus $^{56}$Ni model, respectively.

\clearpage

\clearpage

\end{document}